# NuSTORM μ Ring – Design and Injection Optimization

**David Neuffer[1], Ao Liu[2] and A. Bross**

*Fermilab*
*PO Box 500, Batavia, IL 60510, USA*
*E-mail:* neuffer@fnal.gov

[2] *also Department of Physics, Indiana University*
*Bloomington, IN 47405, USA*

The design of the NuSTORM muon production beam line and storage ring is discussed. The facility includes a pion production target station with a pion collection horn and transport into a straight section of a storage ring. π decay within that straight section provides μ's that are stored within the ring for subsequent decay providing precision electron and muon neutrino beams. The ring and transport designs are described and optimized. Genetic Algorithm optimization of the horn and transport into the ring has been performed, providing a significant increase in intensity.

---

[1]
      Speaker

## 1.Introduction

The nuSTORM project at Fermilab includes proton extraction from the main injector, pion production, collection and injection into a racetrack ring, muon capture from pi decay with $\pi$ and $\mu$ decay in the straight section providing neutrino beams for neutrino studies at the near detector and far detector.[1] The stochastic injection scheme employed by nuSTORM avoids using a separate pion decay channel and fast kickers. The feasibility of the injection was confirmed by Neuffer and Liu [2]. An overview of the facility is displayed in Figure 1.

The muon decay ring is set at a center momentum of 3.8 GeV/c with ±10% momentum acceptance to produce electron and muon neutrino beams with optimal properties for neutrino studies. The pion injection beamline is designed to capture and transport pions within $P_\pi = P_0$ ±10% GeV/c momentum range, where $P_0$ is the design momentum of the pion beamline. $P_0$ is chosen to optimize muon productivity within the 3.8 ±10% GeV/c range. An optimization sets $P_0$ at ~5 GeV/c, as confirmed by G4Beamline calculations.[3, 4]

The stochastic injection scheme calls for both primary pions and secondary stored muons to be transported in the decay straight of the storage ring. The periodic beta functions in the decay straight FODO cells are designed to match both beams. The critical beamline element in the injection scheme is the Beam Combination Section (BCS) of the ring, which includes a defocusing quadrupole, a bending dipole, and a focusing quadrupole. The BCS serves as the injection chicane that brings the pion beam into the decay straight, and simultaneously accommodates the circulating muon beam.

The pions are produced by primary protons at the target, collected by the magnetic horn, and transmitted to the end of the decay straight until extracted, The $\pi$ transport from the target through the horn to the end of the racetrack ring's decay straight is called the pion beamline (PBL) of nuSTORM.

In the following sections we describe the decay ring design, and the matched PBL design, and discuss its performance. In the final section we describe the recent genetic algorithim (GA) optimization of the horn and PBL transport.

## 2.Decay Ring Overview

The goal of the storage ring is to capture and store a large number of $\mu$'s from $\pi$-decay, in order to obtain intense and well-defined neutrino beams fron decay in a production straight section. Thus the ring must have a large acceptance, matched  to the beam sizes obtained from $\pi$ production and decay.  The ring is designed to have a transverse acceptance of 0.002m-rad (unnormalized) and a $P_\mu = 3.8$ ±10% acceptance. The ring has a "racetrack" shape with elongated straight sections to maximize decays along the $\nu$ production straight. The arcs are designed to maintain the acceptance, while including an arc to straight section that can accommodate $\pi$-injection with $\mu$ recirculation (the BCS).

Parameters of the ring are presented in table 1, and $\beta$-functions around the ring are displayed in figure 3. The 180m straight section consists of 24 FODO cells, with parameters chosen to balance the criteria of maximizing the $\mu$-decay neutrino flux, minimizing the number of cells, and minimizing the beam size. The straight section is dispersion-matched with the BCS by using the combination of a dipole and a defocusing quadrupole, at which the dispersion $D_x =$ 2.2 m. The dispersion along with $D_x$ creates a separation of 50 cm between the 3.8 GeV/c reference μ and the 5 GeV/c injection π. This separation distance was confirmed by both PTC tracking in MADX[5] and G4Beamline particle tracking in the magnetic field. The Twiss functions from the production straight FODO cells to the BCS injection point for a 3.8 GeV/c reference muon beam are displayed in Figure 4.

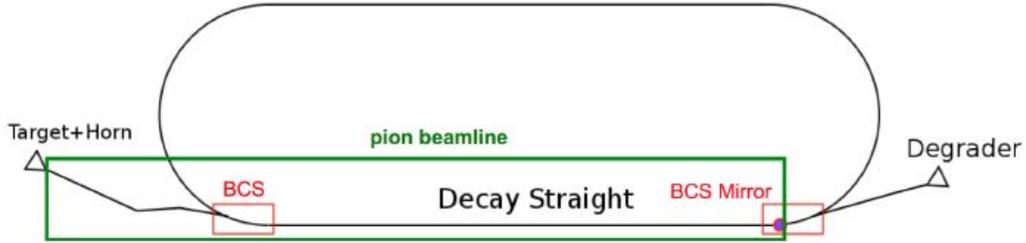

**FIGURE 1.** Schematic overview of the NuSTORM storage ring, highlighting the locations of the pion beam line, the BCS for π/μ the decay straight for π decay for injection and ν production from μ decay.

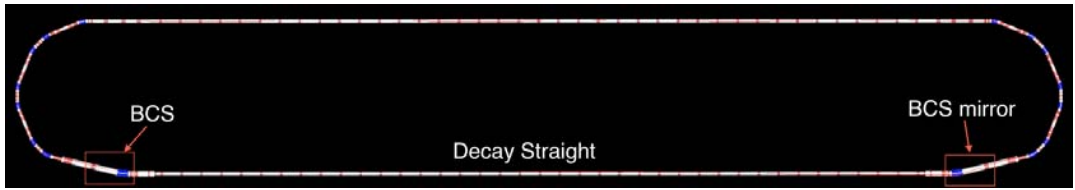

**FIGURE 2.** Scale view of the ring as produced in G4Beamline, (red= quad, blue=dipole, white=drift).

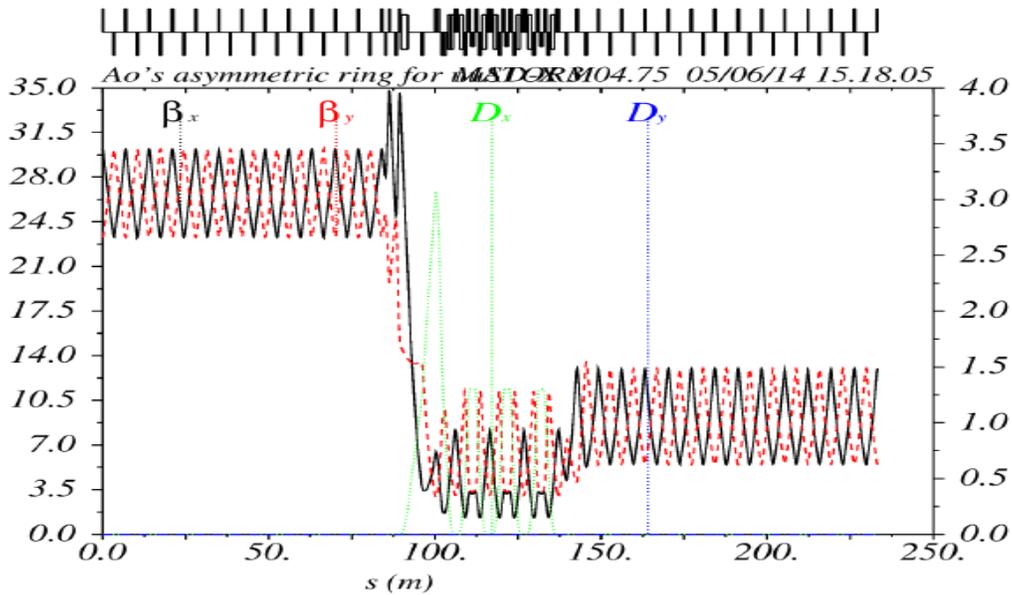

**FIGURE 3.** Betatron functions of a half ring for 3.8 GeV/c μ. (red= $\beta_y$, black=$\beta_x$, green =$D_x$ ). (center of decay straight to center of return straight).

One way to control the beam size in the arc section of the ring is to minimize dispersion. For this, a "double-bend achromat" (DBA) cell structure is used, which has zero dispersion between cells, but non-zero values within each cell. Each arc consists of 3 DBA cells with the BCS insert. The production arc has an enlarged beam size (large $\beta_t$) to minimize ν-beam divergence and maximize π/μ acceptance. The return arc is less constrained and has stronger focusing (smaller $\beta_t$). An interesting feature of the DBA-based lattice is that it is nearly isochronous ($\gamma_t$ =28.5 with $\gamma_\mu$=38) so that the initial bunch structure of the proton beam is maintained within the μ's, improving ring current monitoring diagnostics. The large chromaticity should be corrected; an optimum scheme is not yet developed.

Particle tracking was done with MADX PTC tracking.[5] Starting with a beam that is transversely Gaussian distributed within 2 mm rad, and momentum uniformly distributed within 3.8 10% GeV/c, the beam loss due to the aperture limit is approximately 14% in a single turn, and 40% in 100 turns. 80% of initial μ's decay within 100 turns.

**Table 1:** Decay Ring Parameters

| Parameter | Symbol | Value |
|---|---|---|
| Circumference | C | 468m |
| Tunes | $\nu_x$, $\nu_y$ | 9.72, 7.61 |
| Chromaticity | $\xi_x$, $\xi_y$ | -12.4, -9.2 |
| Long Straight | $L_S$ | 180 m |
| Arc length | $L_L$ | 54m |
| Max β | $\beta_{mas}$ | 34 m |
| Dispersion | $\eta_{max}$, $\eta_{DBA}$ | 3.0, 1.2 m |
| Quad DBA | B', L | 12 T/m, 0.5 m |
| Dipole | B | 4.5T |
| β in straight (3.8 GeV/c μ) | $\beta_{max}$, $\beta_{min}$ | 30, 23 m |
| β in straight (5 GeV/c π) | $\beta_{max}$, $\beta_{min}$ | 38.5/32 m |
| β range in arc | $\beta_{arc}$ | 2/10 |
| β range in return straight | $\beta_{RS}$ | 6/12 |
| Maximum magnet aperture | $a_{mas}$ | 0.3m |

## 3. Injection and Pion Beam Line Design

nuSTORM is designed to use a 100 kW target station. A proton pulse with approximately $10^{13}$ protons at 120 GeV and pulse length of 1.6 s will be extracted from the Fermilab Main Injector and bombarded onto a solid target. The target is partially inserted into a NuMI-like, 300 cm long magnetic horn to collect the pions produced from the target. The horn shape and length are designed to optimize the acceptance of pions within a momentum range of $P_0$ ±10%. The optimum value of $P_0$ was found to be ~5GeV/c. (In simulations π's within ± 18% can produce accepted μ's.)

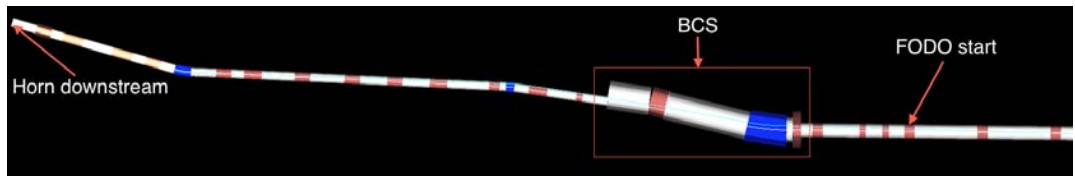

**FIGURE 4.** Scale view of the injection beam line produced in G4Beamline, from horn through entrance into the FODO straight (red= quad, blue=dipole, white=drift).

π's are transported from the target through a chicane into the BCS which places them within the decay straight of the storage ring. The transport and chicane separates the collected π's from primary beam protons and opposite-sign particles while optically matching the beam into the decay straight (see fig. 5). The transport is designed to accept a 0.002m-rad transverse acceptance.

Graphite and Inconel targets were considered and beam produced from 120 GeV protons was tracked through the PBL. For a graphite target, 0.14 π/p are produced at the target within the initial acceptance, resulting in 0.008μ/p within the ring acceptance.[3] With an Inconel target, 0.157 π/p are produced at the target within the initial acceptance, resulting in 0.013μ/p within the ring acceptance. Inconel is preferred because the shorter, denser target results in a denser beam more easily accepted by the ring.

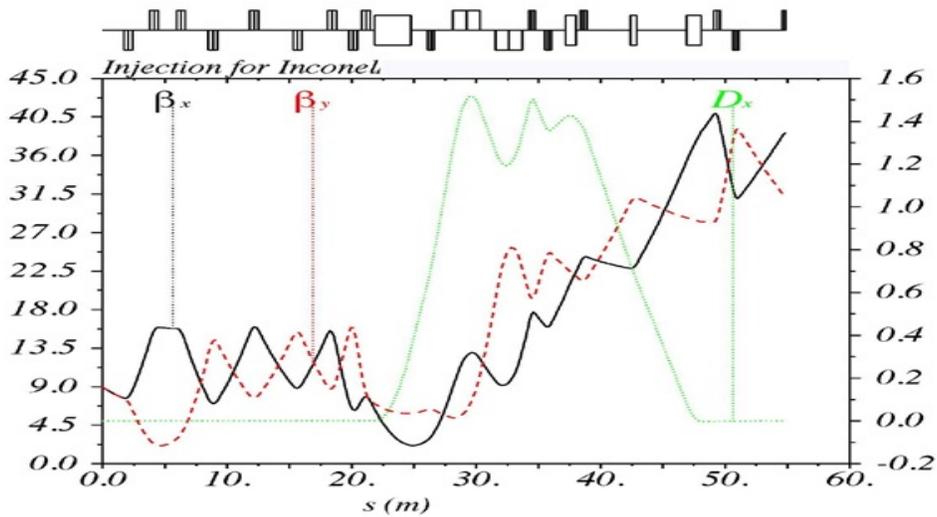

**FIGURE 5.** Betatron functions ($\beta_x$, $\beta_y$, $D_x$) in the injection pion beam line from horn into Decay ring for 5 GeV/c π's.

## 4.Magnetic Horn Optimization

The PBL has many different parameters that can be varied in optimization. In particular the pion collection horn has a large number of parameters that can be varied. The magnetic horn can be characterized by using 7 geometric variables with the horn current and the target location, and these 9 parameters can be varied in optimization.

The Multi-Objective Genetic Algorithm (MOGA) has been widely recognized for its efficiency in searching for extrema for various types of problems, especially when there are more than one optimization objective (goal) and the analytical form for extrema conditions is unknown. MOGA optimization of the horn within the PBL was performed.[6] The 9 horn parameters were set as the genetic variables and the objectives chosen for optimization were the number of π's produced within a 0.002 m-rad acceptance matched into the PBL transport and the number of μ's within 3.8 ± 10% obtained in the decay straight.

In the algorithm, 200 initial cases with different genes are chosen; with best cases chosen for survival. Survivors are combined and mutated; second generation cases are obtained. This process is continued for multiple generations to obtain optima. As each evaluation requires optics rematching and multiparticle calculations, substantial computing power is required. The NERSC multicore computing environment was used.[7]

The algorithm converges within ~66 generations, and results for a 38cm Inconel target are displayed in figure 6. The NUMI horn was used as the basis for the initial design. The reoptimized horn is smaller and requires less current and increases delivered μ's by ~10%, as verified by G4beamline simulation of the optimized system.

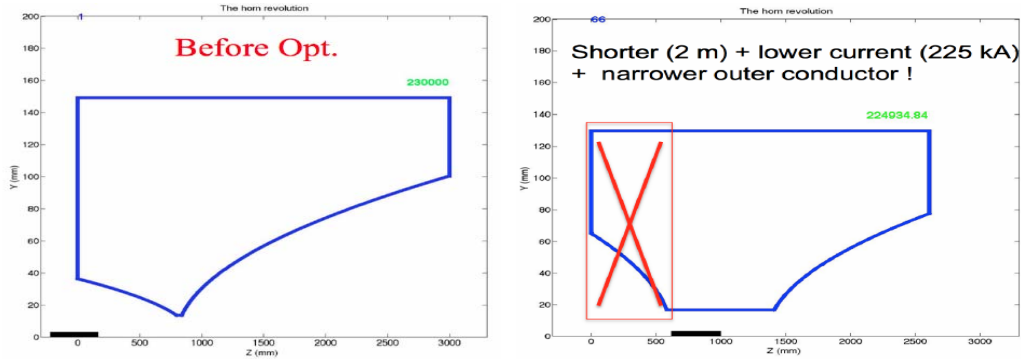

**FIGURE 6.** Horn parameters before and after MOGA optimization of the parameters. The optimum horn configuration is shorter and lower current than the baseline and the production target is moved within the horn. The red X notes the portion of the horn in front of the target; this portion of the horn can be truncated since it cannot affect π motion downstream of production.

## Acknowledgments

We are grateful to the NuSTORM collaboration for their support of this research and to S. Y. Lee for important guidance and assistance. Fermilab is operated by Fermi Research Alliance, LLC under Contract No. De-AC02-07CH11359 with the U. S. Department of Energy.

## References

[1] P. Kyberd et al., *Neutrinos from STORed Muons: Proposal to the Fermilab Physics Advisory Committee* `arXiv:1308.6288`]

[2] D. Neuffer and A. Liu, *Stochastic Injection Scenarios and Performance for NUSTORM*, Proc. IPAC2013, Shanghai, China, TUPFI055, p. 1469 (2013).

[3] A. Liu, A. Bross, D. Neuffer, S. Y. Lee, *NuSTORM Pion Beamline Design Update,* Proc. IPAC2013, Shanghai, China, TUPBA18, p. 562 (2013).

[4] T. Roberts, *G4beamline - A "Swiss Army Knife" for Geant4, optimized for simulating beamlines* (2013), Version 2.12, http://www.muonsinternal.com/muons3/G4beamline

[5] F. Schmidt and H. Grote, *MAD-X and Upgrade from MAD8*, Proc. US Part. Acc. Conf., Portland, Ore USA, p. 3497 (2003). http://mad.home.cern.ch/mad/

[6] A. Liu, A. Bross, D. Neuffer, *NuSTORM Horn Optimization Study,* Proc. IPAC2014, Dresden, Germany, TUPRI005 (2014).

[7] NERSC – National Energy Research Scientific Computing Center, www.nersc,gov/